\documentstyle[aps,multicol,epsfig,prl]{revtex}

\begin{document}
 \input epsf
\draft \preprint{HEP/123-qed}

\title{The anisotropy of granular materials}
\author{F. Alonso-Marroqu\'{\i}n$^{1}$  S. Luding$^{1,2}$ and H. J. Herrmann$^{1}$}
\address{(1) ICA1, University of Stuttgart, 
Pfaffenwaldring 27, 70569 Stuttgart, Germany}
\address{(2) Technische Universiteit Delft (TUD), DelftChemTech, Particle Technology, \\
       Julianalaan 136, 2628 BL Delft, The Netherlands\\}
\date{\today}
\maketitle


\begin{abstract}

The effect of the anisotropy on the elastoplastic response of two dimensional 
packed samples of polygons is investigated here, using molecular dynamics 
simulation.  We show a correlation between fabric coefficients, 
characterizing the anisotropy of the  granular skeleton, and the anisotropy 
of the elastic response. We also study the  anisotropy induced by shearing 
on the subnetwork of the sliding contacts. This anisotropy provides an 
explanation to some features of the plastic deformation of granular 
media. 

\end{abstract}

\begin{multicols}{2}
\section{Introduction}
\label{Intro}

The mechanical behavior of granular materials has been largely 
investigated using constitutive models. These are empirical 
relations which are based on, for example, laboratory tests of soil specimens. 
On the other  hand,  the investigation of the soils at the grain scale,
using discrete element modeling, has become possible in recent years.
These models have provided a valuable  understanding of many 
micro-mechanical aspects of soil deformation.
In particular, numerical simulations of packings of disks  
evidence that  the stress applied on the boundary of the assembly 
is transmitted  through a heterogeneous network of interparticle  
contacts \cite{radjai96}.  The geometric change of this network 
during deformation shows a structural anisotropy induced by 
shearing \cite{thornton86,luding03,laetzel00}. Numerical simulations have also
shown a relevant number of contacts reaching the sliding
condition, even when the sample is isotropically compressed
\cite{alonso04,radjai96}. However, little work has been done in
order to connect the sliding at the contacts to the plastic 
deformation of granular materials.
The aim of this paper is to combine the continuous and the
discrete approaches in the investigation of the anisotropic
response of granular materials, taking into account the 
anisotropy induced in both sliding and non-sliding contacts. 

We will numerically study the elasto-plastic response of
a two-dimensional granular model material. 
The interparticle forces include elasticity, viscous damping and  
friction with the possibility of slippage.
The system is driven by applying stress controlled tests
at the boundary particles. The results show that the traditional
fabric tensor is not sufficient to describe the complex
elasto-plastic response of granular materials. Additional
parameters taking into account the anisotropy of the subnetwork
of the sliding contacts are necessary to include in the 
description of the overall  plastic deformations.

This paper is organized as follows: The details of the
particle model are presented  in Sec. \ref{model}. The contact 
forces are implemented by a Coulomb friction criterion,
and the stress is controlled by the application of suitable
forces at the boundary particles.
The calculation of the constitutive relations is presented 
in Sec. \ref{s-s}. Here we discuss the results in the framework 
of the Drucker-Prager theory of elasto-plasticity.
In Sec. \ref{elastic} we study the effect of the anisotropy 
of the contact network in the incremental elastic response
of the assembly. In Sec. \ref{plastic}   some internal 
variables are introduced in the  constitutive relations. 
These variables take into account the effect of the anisotropy 
induced  in the subnetwork of the sliding contacts on the
plastic deformation of the assembly. 

\section{Model}
\label{model}
 
We present here an extension of those two-dimensional discrete element methods 
which have been used to model granular materials  via polygonal particles 
\cite{tillemans95,kun99}.  The details of the particle generation, the contact 
forces, the  boundary conditions and the molecular dynamics simulations are 
presented in this section.

\subsection{Generation of polygons}
\label{voronoi}

\begin{figure}[t]
  \begin{center}
    \epsfig{file=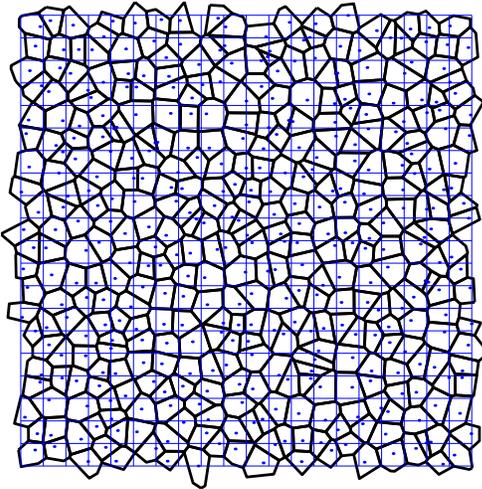,width=0.75\linewidth,clip=1}
    \caption{Voronoi construction used to generate the convex polygons.  
The dots indicate the point used in the tessellation. Periodic boundary 
conditions were used.}
    \label{fig:voronoi}
  \end{center}
\end{figure}

The polygons representing  the particles in this model are generated
by using the method of Voronoi tessellation \cite{kun99}.
This methods is schematically shown in Fig. \ref{fig:voronoi}:
First, a regular square lattice of side $\ell$ is created. Then, we
choose a random point in each  cell of the rectangular grid. Then, each 
polygon is constructed assigning  to  each point that part of the plane 
that is nearer to it than  to any other point. The details of the 
construction of the Voronoi cells can be found in the literature 
\cite{moukarzel92,okabe92}.

Using the Euler theorem, it has been shown analytically that the mean
number of edges of this Voronoi construction must be six \cite{okabe92}. 
The number of edges  of the polygons is distributed between  $4$ and  
$8$ for $98.7\%$ of the polygons. 
We  also found that the orientational  distribution of edges is 
isotropic, and the distribution of areas of polygons is  symmetric 
around its mean value $\ell^2$.  The probabilistic  distribution  of 
areas  follows approximately a    Gaussian  distribution with variance  
of $0.36\ell^2$.

\subsection{Contact forces}
\label{contact_forces}

In order to give a three-dimensional picture of this model, one can 
consider the polygons as a collection of prismatic bodies with 
randomly-shaped polygonal basis. We assume that all the bodies 
have the same thickness $L$. The force between two polygons is 
written as  $\vec{F}= L\vec{f} $ and the mass of the polygons is 
$M=L m $.  

In reality, when two elastic bodies come into contact, 
they have a slight deformation in the contact region. In the 
calculation of the contact force we suppose that the polygons 
are rigid, but we allow them to overlap. Then, we calculate 
the force from this virtual overlap.

\begin{figure}
  \begin{center}
     \epsfig{file=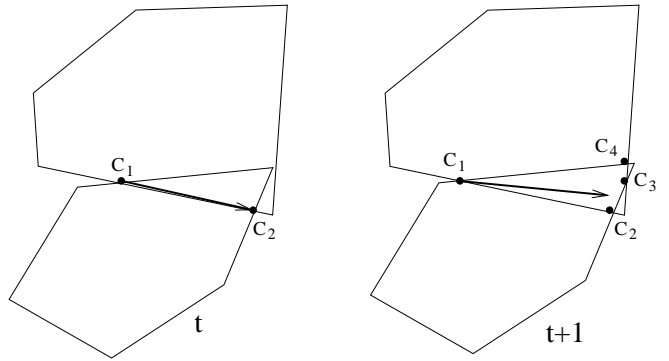,width =1.\linewidth}
   \end{center}
   \caption{ Contact points $C_i$ before (left) and after the 
formation of a pathological contact (right). The vector denotes 
the contact line. $t$ represents the time step.}
    \label{fig:overlap}
\end{figure}

The first step for the calculation of the contact force is the 
definition of the line representing the flattened contact surface 
between the two bodies in contact.  This is defined from the 
contact points resulting from the intersection of the edges of 
the overlapping polygons. In most  cases, we have two contact 
points, as shown in the left of Fig. \ref{fig:overlap}. In such a case, 
the contact line is defined by the vector $\vec{C}=\overrightarrow{C_1 C_2}$ 
connecting these two intersection points. In some pathological
cases, the intersection of the polygons leads to four or six contact 
points. In these cases, we define the contact line by the vector 
$\vec{C}=\overrightarrow{C_1 C_2}+\overrightarrow{C_3 C_4}$ or 
$\vec{C}=\overrightarrow{C_1 C_2}+\overrightarrow{C_3 C_4}+
\overrightarrow{C_5 C_6}$, respectively. 
This choice guarantees a continuous change of the contact line, 
and therefore of the contact forces, during the evolution of the 
contact.

The contact force is separated as $\vec{f}^c=\vec{f}^e+\vec{f}^v$, 
where $\vec{f}^e$  and  $\vec{f}^v$ are the elastic and viscous 
contribution.  The elastic part of the contact force is decomposed 
as $\vec{f^e}= f^e_n \hat{n}^c + f^e_t \hat{t}^c$. The calculation 
of these components is explained below. The unit tangential vector 
is defined as  $\hat{t}^c=\vec{C}/|\vec{C}|$, 
and the normal unit vector $\hat{n}^c$ is taken perpendicular 
to $\vec{C}$. The point of application of the contact force is 
taken as the center of mass of the overlapping polygons.

As opposed to the Hertz theory for round contacts, there 
is no exact way to calculate the normal force between interacting 
polygons. An alternative method has been proposed in order to model 
this force\cite{tillemans95}. In this method, normal elastic force 
is calculated as $f^e_n= -k_n A/L_c$ where $k_n$ is the normal stiffness,  
$A$ is the overlapping area and $L_c$ is a characteristic length 
of the polygon pair. Our choice is $L_c=|\vec{C}|$. 
This normalization is necessary to reflect the fact that the 
normal elastic force must be proportional to an {\it overlapping length}
as it was shown in bidimensional Hertzian contacts \cite{laetzel02}. 

In order to model the quasi-static friction force, we calculate 
the elastic tangential force using an extension of the method 
proposed by Cundall-Strack \cite{cundall79}. An elastic force   
$f^e_t= -k_t \Delta x_t $ proportional to the elastic displacement 
is included at each contact. $k_t$ is the tangential stiffness. 
The elastic displacement $\Delta x_t $ is calculated as the time 
integral of the tangential velocity of the contact during the 
time where the elastic condition  $|f^e_t|<\mu f^e_n$ is satisfied. 
The sliding condition is imposed, keeping this force constant when 
$|f^e_t|=\mu f^e_n$. The straightforward calculation of this elastic 
displacement is given by the time integral starting at the beginning 
of the contact:

\begin{equation}
\Delta x^e_t=\int_{0}^{t}v^c_t(t')\Theta(\mu f^e_n-|f^e_t|)dt',
\label{friction} 
\end{equation}

\noindent
where $\Theta$ is the Heaviside step function and $\vec{v}^c_t$ 
denotes the tangential component of the relative velocity $\vec{v}^c$ 
at the contact.

\begin{equation}
\vec{v}^c=\vec{v}_{i}-\vec{v}_{j}-\vec{\omega}_{i}\times\vec{\ell}_{i}
+\vec{\omega}_{j}\times\vec{\ell}_{j}.
\end{equation}

\noindent
Here $\vec{v}_i$ is the linear velocity and $\vec{\omega}_i$ is the 
angular velocity  of the particles in contact. The branch vector 
$\vec{\ell}_i$ connects the center of mass of particle $i$ with 
the point of application of the contact force. Eq. (\ref{friction})
allows to keep $\Delta x_t$ at a length such that $|f^e_t|$ agrees with
$\mu f^e_n$ during the sliding condition.

Damping forces are included in order to allow for rapid relaxation 
during the preparation of the sample, and to reduce the acoustic 
waves produced during the loading. These forces are calculated as 

\begin{equation}
\vec{f}^c_v  = -m(\gamma_n v^c_n \hat{n}^c + \gamma_t v^c_t \hat{t}^c),
\label{dm0}
\end{equation}

\noindent
with $m=(1/m_i+1/m_j)^{-1}$, the effective mass of the polygons 
in contact. $\hat{n}^c$ and $\hat{t}^c$ are the normal and tangential 
unit vectors defined before, and $\gamma_n$ and $\gamma_t$ are the 
coefficients of viscosity.  These forces introduce time dependent effects
during the cyclic loading. However, we will show that these effects can 
be arbitrarily reduced by increasing the loading time, as it 
corresponds to the quasi-static approximation.

\subsection{Molecular dynamics simulation}
\label{MD}

The evolution of the position $\vec{x}_i$ and the orientation 
$\varphi_i$ of the polygon $i$ is governed 
by the equations of motion:

\begin{eqnarray}
 m_i\ddot{\vec{x}}_i &=&\sum_{c}\vec{f^c_i}  
+\sum_{b}\vec{f}^b_i, \nonumber\\
I_i\ddot{\varphi}_{i} &=&\sum_{c}\vec{\ell}^c_i\times\vec{f^c_i}
+\sum_{b}\vec{\ell}^b_i\times\vec{f}^b_i. 
\label{dm}
\end{eqnarray}

\noindent
Here $m_i$ and $I_i$ are the mass and moment of inertia of the polygon $i$. 
The first summation goes over all particles in contact with this polygon.
According to the previous section, these  forces  $\vec{f^c}$ are given by

\begin{eqnarray}
\vec{f^c}&=&-(k_n A/L_c + \gamma m v^c_n)\vec{n}^c 
-(\Delta x^c_t + \gamma_n m v^c_t)\vec{t}^c, \nonumber \\
\label{dm2} 
\end{eqnarray}
 
\noindent


The second summation on the left hand side of Eq. (\ref{dm}) goes over all the 
edges of the polygons in contact with the  external contour of the 
assembly. We apply a force

\begin{equation}
\vec f^b=-\sigma_1\Delta x^b_2 \hat{x_1} + \sigma_3 \Delta x^b_1 \hat{x}_2  
    -m_i \gamma_b \vec{v_i}.
\label{fbound}
\end{equation} 
 
\noindent
on each selected segment  
$\vec{T}^b=\Delta x^b_1 \hat{x}_1+\Delta x^b_2 \hat{x}_2$ of 
the external contour of the assembly. Here $\hat{x}_1$ and $\hat{x}_2$ are 
the unit vectors of the Cartesian coordinate system. 
$\sigma_1$ and $\sigma_3$ are the components of the stress
we want to apply on the sample, as we see in Subsec. \ref{s-s}.
$m_i$ and $\vec v_i$ are the mass and the velocity of the 
particle $i$ belonging to the boundary. $\vec\ell^b$ is the
vector connecting the center of mass of the boundary particle
$i$ to the center of the boundary segment $\vec T^b$.
 
We use a fifth-order Gear predictor-corrector method for solving 
the equation of motion \cite{allen87}. This algorithm consists of 
three steps. The first step predicts position and velocity of
the particles by means of a Taylor expansion. The second step 
calculates the forces as a function of the predicted positions 
and velocities. The third step corrects the positions and
 velocities in order to optimize the stability of the algorithm. 
This method is much more efficient than the simple Euler approach 
or the Runge-Kutta method, especially for problems where very high 
accuracy is a requirement.
 
The parameters of the molecular dynamics simulations were adjusted 
according to the following criteria: 
1) guarantee the stability of the numerical solution,
2) optimize the time of the calculation, and
3) provide a reasonable agreement with the experimental data.
 
There are many parameters in the molecular dynamics algorithm. 
Before choosing them, it is convenient to make a dimensional 
analysis. In this way, we can keep the scale invariance of the model 
and reduce the parameters to a minimum of dimensionless constants.
First, we   introduce the following characteristic times of the 
simulations:  the loading time $t_0$, the relaxation times  
$t_n=1/\gamma_n$, $t_t=1/\gamma_t$, $t_b=1/\gamma_b$ and the 
characteristic period of oscillation  $t_s=\sqrt{k_n/\rho\ell^2}$ 
of the normal contact.

Using the Buckingham Pi theorem \cite{buckingham14}, one can show that the 
strain response, or any other dimensionless variable measuring the response 
of the assembly during loading, depends only on the following dimensionless 
parameters:  $\alpha_1 = t_n/t_s$, $\alpha_2 = t_t/t_s$,  
$\alpha_3 = t_b/t_s$, $\alpha_4= t_0/t_s$, the ratio $\zeta=k_t/k_n$ 
between the stiffnesses, the friction  coefficient $\mu$ and the ratio 
$\sigma_i/k_n$ between the stresses and the normal stiffness.  

The variables $\alpha_i$ will be called {\it control parameters}. 
They are chosen in order to satisfy the quasi-static approximation, 
i.e. the response of the system does not depend on the loading time, 
but a change of these parameters within this limit does not affect the 
strain response. $\alpha_2 = 0.1$ and $\alpha_2=0.5$ were taken large 
enough to have a high dissipation, but not too large to keep
the numerical stability of the method.  $\alpha_3 = 0.001$ is chosen 
by optimizing the time of consolidation of the sample during the application
of the confining pressure. The ratio $\alpha_4=t_0/t_s=10000 $
was chosen large enough in order to avoid rate-dependence in the strain response, 
corresponding to the quasi-static approximation.  Technically, this is performed by 
looking for the value of  $\alpha_4$ such that a reduction of it by half makes a 
change of the stress-strain relation less than $5\%$.

The parameters $\zeta$ and $\mu$ can be considered as {\it material parameters}. 
They determine the constitutive response of the system, so they must be adjusted 
with the experimental data. In this study, we have adjusted them by comparing the 
simulation of biaxial tests of dense polygonal packings with the corresponding 
tests with dense Hostun sand \cite{marcher01}. First, the initial Young modulus 
of the material is linearly related to the value of normal stiffness of the contact. 
Thus, $k_n=160MPa$ is chosen by fitting the initial slope of the stress-strain 
relation in the biaxial test. Then, the Poisson ratio depends on the ratio 
$\zeta=k_t/k_n$. Our choice $k_t=52.8MPa$ gives an initial Poisson ratio of $0.07$. 
This is obtained from the initial slope of the curve of volumetric strain versus 
axial strain. The angles of friction and the dilatancy are increasing functions 
of the friction coefficient $\mu$. Taking $\mu=0.25$ yields angles of friction 
of $30-40$ degrees and dilatancy angles of $20-30$ degrees.  The experimental data 
yields angles of friction between $40-45$ degrees and
dilatancy angles between $7-14$ degrees. A better adjustment would be made by 
including different void ratios in the simulations, but this is out of the scope 
of this work.

\section{stress-strain relation}
\label{s-s}

The characterization of the macroscopic state of a granular material in static equilibrium
is usually given by the Cauchy stress tensor. The average of this tensor over
the assembly leads to \cite{bagi99}

\begin{equation}
\sigma_{ij} = \frac{1}{A}\sum_{b} x^b_{i} f^b_{j}
\label{cauchy}
\end{equation}

The sum goes over all the forces acting over the boundary of the assembly.
$\vec{x}^b$ is the point of application of the boundary force $\vec{f}^b$. This force
is defined in Eq. (\ref{fbound}). $A$ is the area enclosed by the boundary. The sum goes 
over all the boundary forces of the sample.  Inserting Eq. (\ref{fbound}) in 
Eq. (\ref{cauchy}) and taking the equilibrium condition $\vec v_i =0$ leads to

\begin{equation}
\sigma= \frac{1}{A}\left[ \begin{array}{cc}
                  -\sigma_1\sum_b{x_b\Delta y_b} &  \sigma_3\sum_b{x_b\Delta x_b}   \\
                  -\sigma_1\sum_b{y_b\Delta y_b}  & \sigma_3\sum_b{y_b\Delta x_b}   
                 \end{array} \right].
\end{equation}

\noindent
Those sums can be converted into integrals over closed loops and the calculation of such 
integrals leads to

\begin{equation}
   \sigma        =  \left[ \begin{array}{cc}
                    \sigma_1& 0  \\
                    0  & \sigma_3  
                     \end{array} \right].
\label{stress1}
\end{equation}

\noindent
Thus, the stress controlled test is restricted to stress states without off-diagonal
components.  So we can simplify the notation introducing the {\it pressure} $p$ and the 
{\it deviatoric stress} $q$ in the components of the {\it stress vector}

\begin{equation}
\tilde{\sigma}=\left[ \begin{array}{c}  p\\q \end{array} \right]  
= \frac{1}{2} \left[ \begin{array}{c} \sigma_1+\sigma_3 \\
                                      \sigma_1-\sigma_3  \end{array} \right].
\label{stv}
\end{equation}

In the same way, the incremental strain tensor can be calculated from  the average of the 
displacement gradient over the area of the RVE. It has been shown \cite{bagi96} that it 
can be  transformed in a sum over the boundary of the sample  

\begin{equation}
d \epsilon _{ij} = \frac{1}{2 A}\sum_{b} {(\Delta u^b_{i} N^b_{j} + \Delta u^b_{j} N^b_{i})}
\end{equation}

Here $\Delta \vec{u}^b$ is the displacement of the boundary segment, that is calculated
from the linear displacement $\Delta\vec{x}$ and the angular rotation 
$\Delta\vec{\phi}$ of the 
polygons belonging to it, according to

\begin{equation}
\Delta \vec{u}^b = \Delta \vec{x} + \Delta \vec{\phi} \times \vec{\ell}. 
\end{equation}

From the eigenvalues  $d\epsilon_1$ $d\epsilon_3$ of  
$d \epsilon_{ij}$ we define the  {\it volumetric } and {\it deviatoric}  
components of the strain as the components of the 
{\it incremental strain vector}:

\begin{equation}
d\tilde{\epsilon}=\left[ \begin{array}{c} de\\d\gamma \end{array} \right]
  = -\left[ \begin{array}{c} d\epsilon_1+d\epsilon_3 \\
                                       d\epsilon_1-d\epsilon_3  \end{array} \right].     
\label{strain}
\end{equation}

By convention $de>0$ corresponds to a compression of the sample. 
We are going to assume a rate-independent constitutive relation between the 
incremental stress and incremental strain tensor. In this case the incremental
relation can  generally be written as \cite{darve95}:

\begin{equation}
d\tilde{\epsilon}=M(\hat{\theta},\tilde{\sigma})d\tilde{\sigma},
\label{dce}
\end{equation}

\noindent
where $\hat{\theta}$ is the unitary vector defining a specific direction in 
the stress space:

\begin{equation}
\hat{\theta}= \frac{d\tilde{\sigma}}{|d\tilde{\sigma}|}
\equiv \left[ \begin{array}{c}  \cos \theta\\ \sin \theta \end{array} \right],
 ~~|d\tilde{\sigma}|=\sqrt{dp^2+dq^2}.
\label{dir}
\end{equation}

The constitutive relation results from the calculation of 
$d\tilde{\epsilon}(\theta)$, where each value of $\theta$ is related to a 
particular mode of loading. Some special modes are listed in Table I.

In order to compare the resulting incremental response to the elasto-plastic 
theory, it is necessary to assume that the incremental strain can be separated 
into an elastic (recoverable) and a plastic (unrecoverable) component:

\begin{equation}
d\tilde{\epsilon}  = d\tilde{\epsilon}^e+ d\tilde{\epsilon}^p,
\label{elastoplastic}
\end{equation}

\begin{equation}
d\tilde{\epsilon}^e= D^{-1}(\tilde{\sigma})d\tilde{\sigma},
\label{eq:elastic}
\end{equation}

\begin{equation}
d\tilde{\epsilon}^p= J(\theta,\tilde{\sigma})d\tilde{\sigma}.
\label{eq:plastic}
\end{equation}

\begin{table}[b]
\begin{center}
\begin{tabular}{|r|c|rr|}
\hline
$ 0^o    $ & isotropic compression & $dp>0 $       & $dq=0$        \\         
$ 45^o   $ & axial loading          & $d\sigma_1>0$& $d\sigma_3=0$ \\         
$ 90^o   $ & pure shear             & $dp=0$       & $dq>0$        \\  
$ 135^o  $ & lateral loading        & $d\sigma_1=0$& $d\sigma_3>0$ \\
$ 180^o  $ & isotropic expansion    & $dp<0$       & $dq=0$        \\
$ 225^o  $ & axial stretching       & $d\sigma_1<0$& $d\sigma_3=0$ \\         
$ 270^o  $ & pure shear             & $dp=0$       & $dq<0$        \\     
$ 315^o  $ & lateral stretching     & $d\sigma_1=0$& $d\sigma_3<0$ \\  
\end{tabular}
\label{tab:modes}
\caption{Principal modes of loading according to the orientation of $\hat{\theta}$}
\end{center}
\end{table}

Here, $D^{-1}$ is the inverse of the stiffness tensor $D$, and $J=M-D^{-1}$ the 
flow rule of  plasticity~\cite{vermeer84}. 
They can be obtained from the calculation of  $d\tilde{\epsilon}^e(\theta)$ 
and $d\tilde{\epsilon}^p(\theta)$ as we will see below.

The method presented here to calculate the strain response has 
been used  on sand experiments \cite{poorooshasb67}. 
It was introduced by Bardet  \cite{bardet94b} in the calculation of the 
incremental response using discrete element methods. 
This method will be used to determine the elastic 
$d\tilde{\epsilon}^e$ and plastic $d\tilde{\epsilon}^p$ components of the 
strain as function of the stress state $\tilde{\sigma}$ and the stress 
direction $\hat\theta$.
  
First, it is isotropically compressed until it reaches the stress value 
$\sigma_1=\sigma_3=p-q$. Next, it is subjected to axial loading in order 
to  increase the axial stress $\sigma_1$ to $p+q$. When the stress state 
with  pressure $p$ and deviatoric stress $q$ is reached, the sample is 
allowed to relax. 
Loading the sample from $\tilde{\sigma}$  to  $\tilde{\sigma}+d\tilde{\sigma}$   
the strain increment $d\tilde{\epsilon}$ is obtained. 
Then the sample is unloaded to $\tilde{\sigma}$ and one finds a 
remaining strain  $d\tilde{\epsilon}^p$,  that corresponds to the plastic  
component of the incremental  strain. For small stress increments the 
unloaded stress-strain path is almost elastic. Thus, the difference   
$d\tilde{\epsilon}^e = d\tilde{\epsilon}- d\tilde{\epsilon}^p$ can 
be taken as the elastic component of the strain.   This procedure is 
implemented on different "clones" of the same  sample, choosing different 
stress directions and the same stress amplitude in each one of them.

This method is based on the assumption that the strain response after a reversal 
loading is completely elastic. However, numerical simulations have shown that 
this assumption is not strictly true, because sliding contacts are always observed
during the unload path \cite{calvetti02,alonso04}. In order to 
overcome this difficulty, Calvetti et al. \cite{calvetti02} calculate the 
elastic part by removing the frictional condition  setting $\mu=\infty$, 
and measuring the purely elastic response $d\tilde{\epsilon}^{ns}$ of the 
assembly. Then the plastic component of the strain can be calculated as
$d\tilde{\epsilon}^p = d\tilde{\epsilon}- d\tilde{\epsilon}^{ns}$. 

In our simulations, we have observed that the plastic deformation during 
the reversal stress path is less than $1\%$ of the corresponding value of the 
elastic response. Within this margin of error, the method proposed by 
Bardet can be taken as a reasonable approximation to describe the 
elasto-plastic response. 

\begin{figure}[t]
  \begin{center}
    \epsfig{file=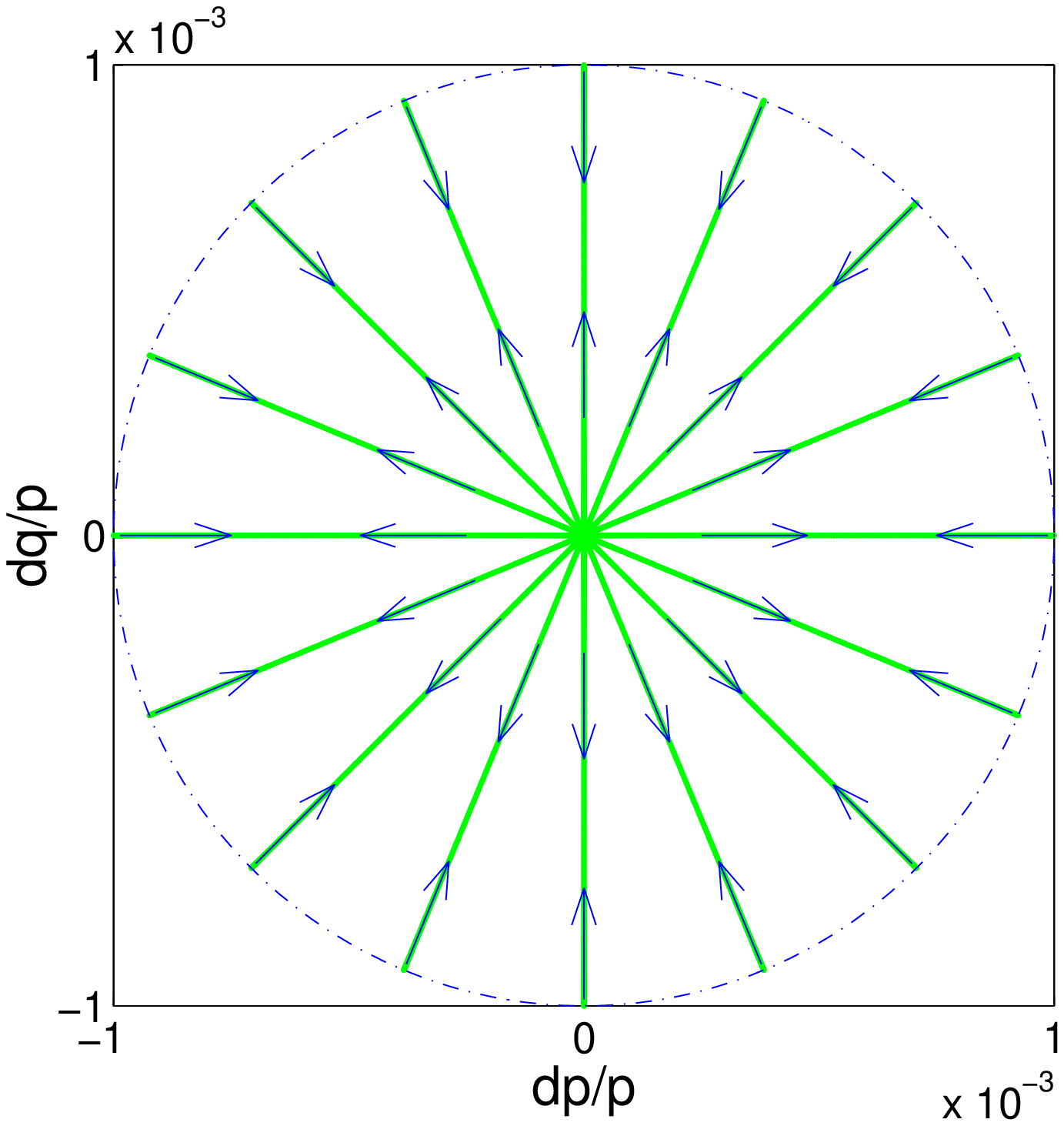,width=.75\linewidth,clip=1}
    \epsfig{file=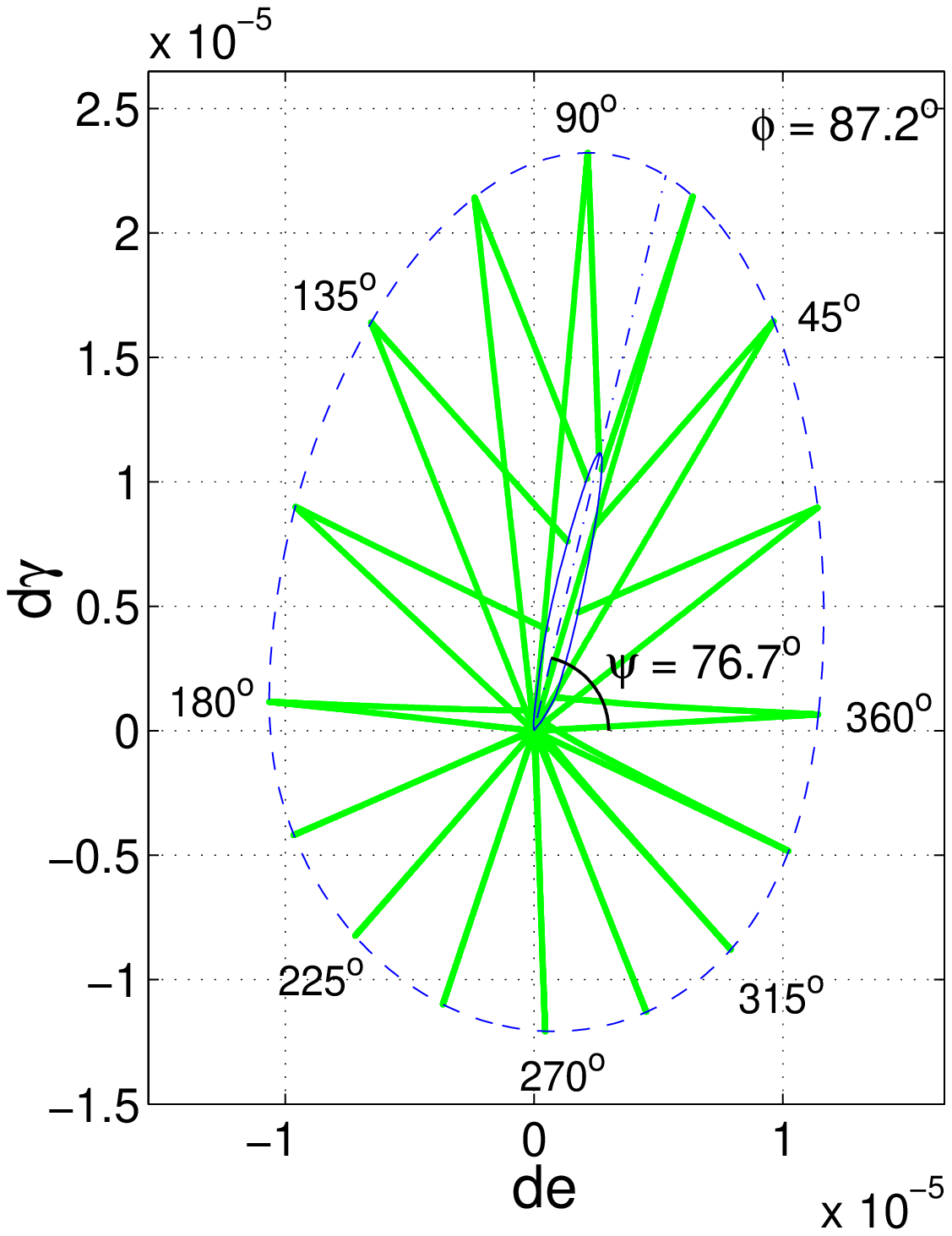,width=.8\linewidth,clip=1}
    \caption{Stress - strain relation resulting from the load - unload test. 
Dotted lines represent the paths in the stress and strain spaces. 
The dash-dot line shows the strain envelope response and the solid 
line is the plastic envelope response.}
    \label{fig:envelopes}
  \end{center}
\end{figure}

Fig. \ref{fig:envelopes} shows the load-unload stress paths and the 
corresponding strain response when an initial stress state with 
$\sigma_1=200kPa$ and  $\sigma_3=120kPa$ is chosen. The end of the 
load paths in the stress space maps into a strain envelope response 
$d\tilde{\epsilon}(\theta)$ in the strain space. Likewise, the end 
of the unload paths maps into a plastic envelope response 
$d\tilde{\epsilon}^p(\theta)$. The {\it yield direction} $\phi$ can 
be found from this response, as the direction in the stress space where 
the plastic response is maximal. In this example, this is around 
$\theta=87.2^o$. The {\it flow direction} $\psi$ is given by the 
direction of the maximal plastic response in the strain space, 
which is around  $76.7^o$. The fact that these directions do not 
agree reflects a {\it non-associated flow rule}, as it is observed 
in experiments on realistic soils \cite{poorooshasb67}.

\section{Elastic response}
\label{elastic}

Hooke's law of elasticity states that the stiffness tensor of isotropic materials 
can be written in terms of two material parameters, i.e. the Young modulus $E$ 
and the Poisson ratio $\nu$ \cite{landau86}  
However, the isotropy is not fulfilled when the sample is subjected to deviatoric 
loading.  Indeed, numerical simulations \cite{thornton86,cundall82} and  
photo-elastic experiments \cite{drescher72} on granular materials 
show that loading induces a significant deviation from isotropy in the 
contact  network.

\subsection{Anisotropy of the contact network}

\begin{figure}
 \begin{center}
 \epsfig{file=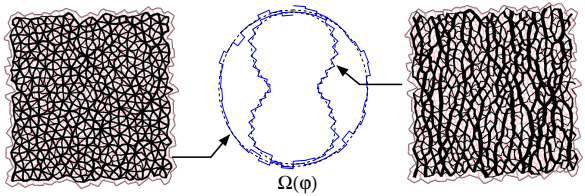,width=\linewidth}
 \end{center}
 \caption{The lines show the branch vectors for $\sigma_1=\sigma_3=160kPa$ (left) 
and $\sigma_1=272kPa$ and $\sigma_3=48kPa$ (right). 
The orientational distribution of branch vectors is shown for both cases.}
 \label{fig:branch}
\end{figure}

The anisotropy of the granular sample can be characterized by the 
distribution of the orientations of the branch vectors $\vec{\ell}$, 
as shown in Fig. \ref{fig:branch}. Each branch vector connects the 
center of mass of the polygon to the center of application of the 
contact force. Fig. \ref{fig:branch} shows the branch vectors for 
two different stages of loading.  The structural changes of 
micro-contacts are principally due to the opening of contacts 
whose branch vectors are oriented nearly perpendicular to the 
loading direction. Let us call $\Omega(\varphi)\Delta\varphi$ 
the number of contacts per particle whose branch vector is 
oriented between the angles $\varphi$ and  
$\varphi + \Delta\varphi$, measured with respect to the direction 
along which the sample is loaded. The anisotropy of the contact 
network can be accurately described by a truncated 
series expansion.

\begin{equation}
\Omega(\varphi) \approx \frac{N_0}{2\pi}\big[a_0 + a_1\cos(2\varphi) + a_2\cos(4\varphi) \big].
\label{eq:fourier1}
\end{equation}

\begin{figure}[b]
 \begin{center}
 \epsfig{file=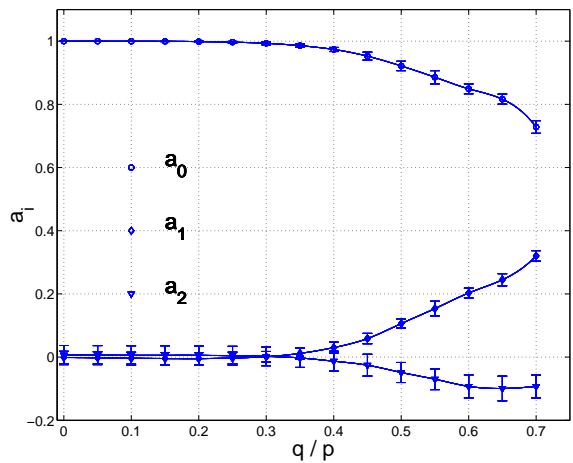,width=0.7\linewidth,angle=-90}
 \end{center}
 \caption{Fabric coefficients of the distribution of the contact normal vectors.
          They are defined in Eq. (\ref{eq:fourier1}).}
 \label{fig:fourier1}
\end{figure}

Here $N=N_0 a_0$ is the average coordination number of the polygons, 
whose initial value $N_0 = 6.0 $ reduces as the load is increased.  
The parameters $a_0$, $a_1$ and $a_2$ can be related respectively 
to the zeroth, second and fourth order fabric tensor defined in other 
works to characterize the orientational distribution of the contacts
\cite{thornton86,laetzel02}. Here, they will be called 
{\it fabric coefficients}. The dependence of the fabric coefficients 
on the stress ratio $q/p$ is shown in Fig. \ref{fig:fourier1}.  
We observe that for stress states satisfying $q<0.4p$ there are 
almost no open contacts. Above this limit a significant number of
contacts are open, leading to an anisotropy in the contact network. 
Fourth order terms in the Fourier expansion are necessary in order 
to accurately describe this distribution.

\subsection{Anisotropic stiffness}

We will investigate the effect of the anisotropy of the contact network
on the stiffness of the material. The most general linear relation between 
the incremental stress and the incremental elastic strain for anisotropic materials 
is given by

\begin{equation}
  \label{eq:elastic tensor}
  d\sigma_{ij} = D_{ijkl} d\epsilon^e_{kl},
\end{equation}

\noindent
where $D_{ijkl}$ is the stiffness tensor \cite{landau86}. Since the stress 
and the strain are symmetric tensors, one can reduce their number of 
components from $9$ to $6$, and the number of components of the stiffness 
tensor from $81$ to $36$. Further, by transposing Eq. (\ref{eq:elastic tensor}) 
one obtains that $D_{ijkl}=D_{jilk}$, which reduces the constants from $36$ 
to $21$. In the particular case of isotropic materials, it has been shown 
that the number of constants can be reduced to $2$ \cite{landau86}:

\begin{equation}
  \label{eq:hooke law}
  d\epsilon^e_{ij} = \frac{1}{E}[(1-\nu)d\sigma_{ij} - v \delta_{ij}d\sigma_{kk}].
\end{equation}

Here $E$ is the Young modulus and $\nu$ the Poisson ratio. The stress-strain
relation of Eq. (\ref{eq:elastic tensor}) has been inverted to compare to the
elasto-plastic relation of Eq. (\ref{eq:elastic}).
The description of  the general case of the anisotropic elasticity with 
$21$ constants does not seem trivial. 
However, since we consider here only plane strain deformations, we
can perform further simplifications. We take a coordinate system oriented in the 
principal stress-strain directions. Thus, the only non-zero components are 
$d\sigma_{11} \equiv d\sigma_1$ and $d\sigma_{33} \equiv d\sigma_3$ for the 
stress and $d\epsilon_{11} \equiv d\sigma_1$ and 
$d\epsilon_{33} \equiv d\sigma_3$ for the strain. The anisotropic elastic 
tensor connecting these components contains only three independent parameters. 
We can write Eq. (\ref{eq:elastic tensor}) as

\begin{equation}
\left[ \begin{array}{c}
d\epsilon^e_1 \\
d\epsilon^e_3
\end{array} \right]
=\frac{1}{E}\left[ \begin{array}{cc}
1-\alpha    & -\nu \\
-\nu &  1+\alpha\end{array} \right]
\left[ \begin{array}{c}
d\sigma_1 \\
d\sigma_3
\end{array}\right].
\label{eq:hooke}
\end{equation}

The additional parameter $\alpha$ is included here to take into account the 
anisotropy. When $\alpha=0$, we recover Hooke's law  of 
Eq. (\ref{eq:hooke law}). Eq. (\ref{eq:elastic}) is calculated from Eq. (\ref{eq:hooke}) 
by performing the transformation in the coordinates of the volumetric strain 
$de=-d\epsilon_1-d\epsilon_2$ and deviatoric strain $d\gamma=d\epsilon_2-d\epsilon_1$, 
and  the pressure $p=(\sigma_1+\sigma_2)/2$ and the deviatoric stress  
$q=(\sigma_1-\sigma_2)/2$. One obtains:

\begin{equation}
\left[ \begin{array}{c}
de^e \\
d\gamma^e
\end{array} \right]
=\frac{2}{E}\left[ \begin{array}{cc}
                  1-\nu&   -\alpha  \\
                  -\alpha  &   1+\nu  
\end{array} \right]
\left[ \begin{array}{c}
dp \\
dq
\end{array}\right]
\label{eq:hooke2}
\end{equation}

\noindent

In the isotropic case $\alpha=0$ this matrix is diagonal. The inverse of the 
diagonal terms are the bulk modulus $K=E/2(1-\nu)$ and the shear modulus 
$G=E/2(1+\nu)$. The anisotropy $\alpha\ne0$ couples the compression mode 
with the shear deformation such that the compression of the sample will 
produce deviatoric deformation. This coupling can be observed from the 
inspection of the elastic part of the strain envelope responses  
$d\tilde{\epsilon}^e(\theta)$ as shown in Fig. \ref{fig:elastic envelopes}. 
For stress values such as $q/p\le 0.4$ the stress envelope responses collapse 
on to the same ellipse. This response can be described by Eq. (\ref{eq:hooke2}) 
taking $\alpha=0$. For stress values satisfying  $q/p>0.4$ there is a coupling 
between compression and shear deformations and it is necessary to take 
$\alpha\ne0$ in Eq. (\ref{eq:hooke2}).

\subsection{Stiffness and fabric}

\begin{figure}[t]
 \begin{center}
 \epsfig{file=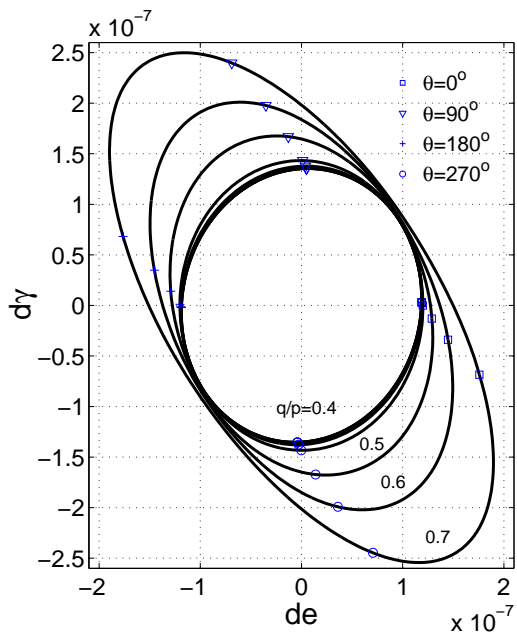,width=0.8\linewidth}
 \end{center}
 \caption{Elastic strain envelope responses $d\tilde\epsilon^e(\theta)$. 
They are calculated for a pressure $p=160kPa$ and taking deviatoric 
stresses with $q=0.0p$ (inner), $0.1p$, ...,$0.7p$ (outer).}
 \label{fig:elastic envelopes}
\end{figure}

Comparing the calculation of the elastic response in Fig. 
\ref{fig:elastic envelopes} with the anisotropy of the contact network  
shown in Fig. \ref{fig:fourier1}, a certain correlation is evident
between the stiffness tensor and the fabric coefficients of
Eq. (\ref{eq:fourier1}). 
We observe that Hooke's law is valid in the regime $q/p<0.4$ where the 
contact network is isotropic. Moreover, we observe that the opening of 
the contacts, whose branch vectors are almost perpendicular to the direction 
of the load, produces a reduction of the stiffness in this direction. 
This results in an anisotropic elasticity.

We are going to find a simple relation between the orientational distribution 
of the contacts and the parameters of the stiffness. These three parameters 
are calculated from the elastic response by the introduction of the quadratic
form of $D^{-1}$:

\begin{equation}
R(\theta) \equiv \hat\sigma^T D^{-1} \hat\sigma =\frac{dp~de^e+dq~ d\gamma^e  }{dp^2+dq^2}.
\label{eq:R}
\end{equation}

\noindent
Here $\hat\sigma^T$ is the transpose of $\hat\sigma$, which is defined in 
Eq. (\ref{dir}).
This function can be directly obtained from the elastic part of the strain 
response $d\tilde\epsilon^e(\theta)$. On the other hand, replacing 
Eq. (\ref{eq:hooke2}) in Eq. (\ref{eq:R}) one can express $R$ in terms of the 
parameters of the stiffness tensor: 

\begin{equation}
R(\theta)= \frac{2}{E}\big[1-\nu\cos(2\theta)-\alpha\sin(2\theta) \big].
\end{equation} 

\noindent

Using this equation, the parameters $E$, $\nu$ and $\alpha$ are evaluated from 
the Fourier coefficients of $R$:

\begin{eqnarray}
\frac{1}{E}& = & \frac{1}{4\pi}\int_0^{2\pi} R(\theta) d\theta,\\
        \nu& = &-\frac{E}{2\pi}\int_0^{2\pi} R(\theta)\cos(2\theta) d\theta,\\
     \alpha& = &-\frac{E}{2\pi}\int_0^{2\pi} R(\theta)\sin(2\theta) d\theta.
\end{eqnarray}  

\begin{figure}[b]
 \begin{center}
 \epsfig{file=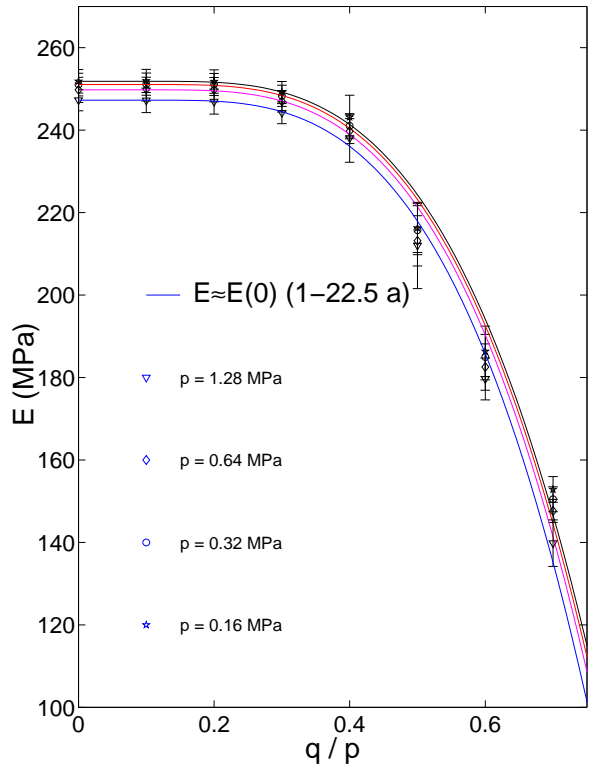,width=0.9\linewidth}
 \end{center}
 \caption{Young modulus. The solid line is the linear approximation of $E(d)$. 
         See Eq. (\ref{tay}).}
 \label{young}
\end{figure}

\begin{figure}[t]
 \begin{center}
 \epsfig{file=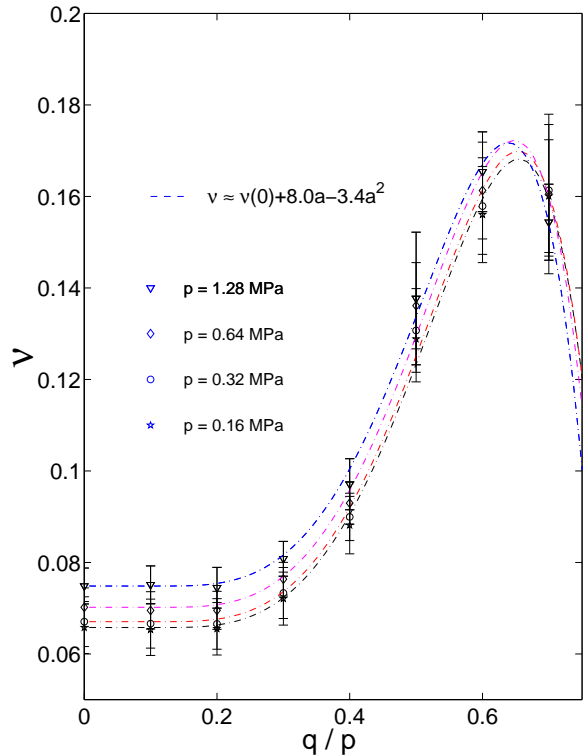,width=0.9\linewidth}
 \end{center}
 \caption{Poisson's ratio. The dashed line is the quadratic approximation 
of $\nu(d)$. See Eq. (\ref{tay}).}
 \label{poisson}
\end{figure}

\begin{figure}[t]
 \begin{center}
 \epsfig{file=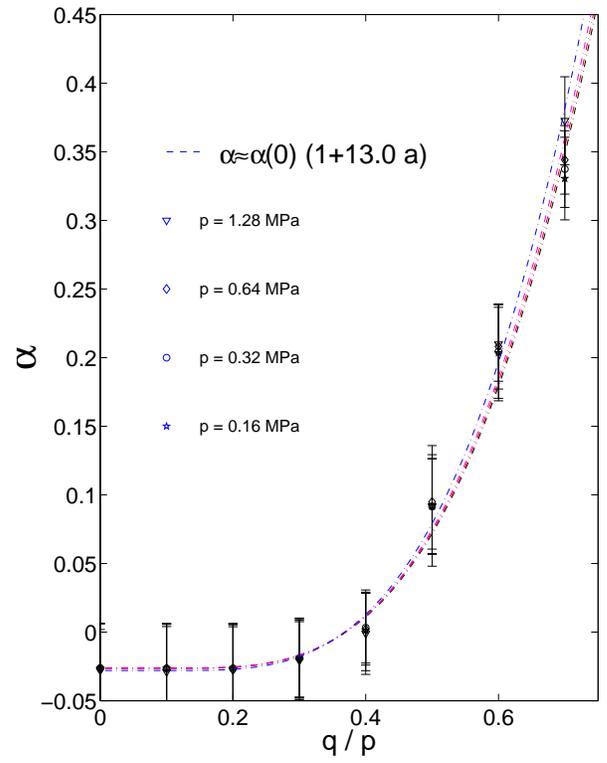,width=0.9\linewidth}
 \end{center}
 \caption{Anisotropy parameter. The dashed line is the linear approximation of 
          $\alpha(d)$. See Eq. (\ref{tay}).}
 \label{aniso}
\end{figure}

Figs. \ref{young}, \ref{poisson} and \ref{aniso} show the results of the 
calculation of the Young modulus $E$, the Poisson ratio $\nu$ and the 
anisotropy factor $\alpha$. Below the limit of isotropy, Hooke's law can be 
applied: $E\approx E_0$, $\nu \approx \nu_0$ and $\alpha \approx 0$. On the 
other hand, above the limit of isotropy a reduction of the Young modulus is 
found, along with an increase of the Poisson ratio and the anisotropy factor. 
In order to evaluate the dependence of these parameters on the fabric 
coefficients $a_i$ from Eq. (\ref{eq:fourier1}), we introduce an internal 
variable measuring the degree of anisotropy. This variable is denoted by 
$a$ and is defined as the percentage change of the total number of contacts.

\begin{equation}
  \label{eq:a}
  a\equiv\frac{N_0-N}{N_0}= 1-a_0,
\end{equation}

\noindent
where $a_0$ is defined in Eq. (\ref{eq:fourier1}). The dependence of the parameters 
of the stiffness tensor on the internal variable $a$ is evaluated by developing 
the Taylor series around $a=0$:

\begin{eqnarray}
E(a)     &=& E(0)+E'(0)a+                O\left(a^2\right), \nonumber \\
\alpha(a)&=& \alpha (0)+\alpha'(0)a+     O\left(a^2\right), \\
\label{tay}
\nu(a)   &=& \nu(0)+\nu'(0)a+\nu''(0)a^2+O\left(a^3\right). \nonumber
\end{eqnarray}

The coefficients of this expansion are calculated from the best fit of those 
expansions. Figs. \ref{young} and \ref{aniso} show that the linear approximation 
is good enough to reproduce the Young modulus and the anisotropy factor. 
The fit of the Poisson ratio, is shown in Fig. \ref{poisson}. The fitting with 
only one internal variable requires the inclusion of a quadratic approximation.  
To obtain more accurate relations, it may be necessary to introduce a more 
complex dependence on the fabric coefficients of Eq. (\ref{eq:fourier1}).

\section{Plastic deformations}
\label{plastic}

In the elasto-plastic models of soils the  plastic deformation is
calculated by introducing a certain number of hypothetical surfaces 
\cite{roscoe68,nova79,dafalias86a,vermeer84}. 
In the Drucker-Prager models, the so-called plastic flow rule is calculated 
from the yield surface and the plastic potential 
\cite{roscoe68,nova79,vermeer84}. In the bounding surface plasticity, 
it is calculated from the loading surface and bounding surfaces 
\cite{dafalias86a} We will see that it is possible to calculate 
the parameters of the flow rule of plasticity, the flow direction, the yield 
direction and the modulus of plasticity, directly from the stress envelope 
response $d\tilde{\epsilon}^p(\theta)$ without introducing such abstract 
surfaces. 

\subsection{Flow rule}
\label{flow rule}

In Fig. \ref{fig:envelopes} we found that the plastic envelope response lies 
almost on a straight line, as is predicted by the Drucker-Prager theory
\cite{drucker52}. 
This motivates us to define the parameters describing the plasticity in the same way 
as this theory: i.e. the yield direction $\phi$, the flow direction $\psi$, 
and the plastic modulus $h$.

The yield direction is given by the incremental stress direction $\phi$ with 
maximal plastic deformation

\begin{equation}
|d\tilde{\epsilon}^p(\phi)|=\max_{\theta}{|d\tilde{\epsilon}^p(\theta)|},
\label{yd}
\end{equation} 

\noindent
The flow direction is defined from the orientation of the plastic response at 
its maximum value

\begin{equation}
\psi = atan2(d\gamma^p,de^p)\left|_{\theta=\phi}\right.     
\label{fd}
\end{equation}

\noindent
Here $atan2(y,x)$ is the four quadrant inverse tangent of the real parts of the
elements of x and y.( $ -\pi <= atan2(y,x) <= \pi$). 
The plastic modulus is  obtained from the modulus of the maximal plastic 
response.

\begin{equation}
\frac{1}{h} = \frac{|d\tilde{\epsilon}^p(\phi)|}{|d\tilde{\sigma}|}.
\label{eq:hardening}
\end{equation} 

\noindent
The incremental plastic response can be expressed in terms of these quantities. 
Let us define the unitary vectors $\hat{\psi}$ and $\hat{\psi}^{\perp}$. The 
first one is oriented in the direction of $\psi$ and the second one is the 
rotation of $\hat{\psi}$ of $90^o$. The plastic strain is written as:

\begin{equation}
d\tilde{\epsilon}^p(\theta) = \frac{1}{h}\left[ \kappa_1(\theta)\hat{\psi} 
+ \kappa_2(\theta)\hat{\psi}^{\perp} \right],
\label{pl1}
\end{equation}

\noindent
where the plastic components $\kappa_1(\theta)$ and $\kappa_2(\theta)$ are given by

\begin{eqnarray}
\kappa_1(\theta) &=&  h (d\tilde{\epsilon}^p \cdot \hat{\psi}) \nonumber\\ 
\kappa_2(\theta) &= & h (d\tilde{\epsilon}^p \cdot \hat{\psi}^{\perp}).
\label{eq:plofiles}
\end{eqnarray}

\begin{figure}[t]
 \begin{center}
 \epsfig{file=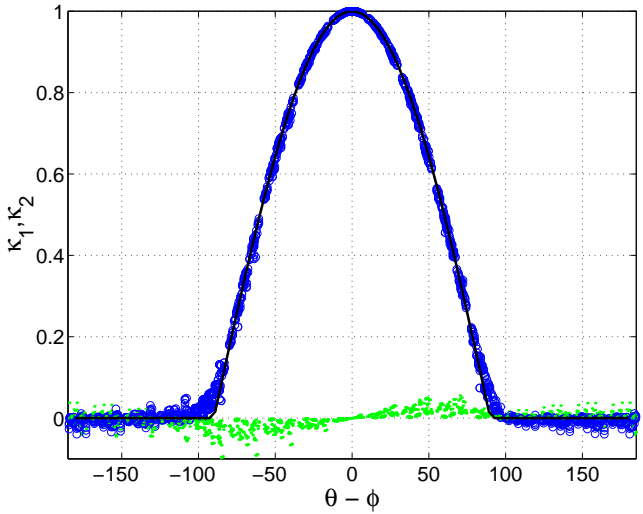,width=\linewidth}
 \end{center}
 \caption{Plastic components $\kappa_1(\theta)$ (circles) and $\kappa_2(\theta)$ (dots). The results for different stress values have been superposed. The solid line represents the truncated cosine function. }
 \label{profile}
\end{figure}

These functions are calculated from the resulting plastic response taking 
different stress values. The results are shown in Fig. \ref{profile}. 
We found that the functions $\kappa_1(\theta-\psi)$ collapse on to the same 
curve for all the stress states. This curve fits well to a cosine function, 
truncated to zero for the negative values. The profile $\kappa_2$ depends 
on the stress ratio we take. This dependency is difficult to evaluate, 
because the values of this function are of the same order as the 
statistical fluctuations. In order to simplify the description of the 
plastic response, the following approximation is made: 

\begin{equation}
\kappa_2(\theta)\ll \kappa_1(\theta)\approx \langle\cos(\theta-\phi)\rangle
                              =\langle\hat{\phi}\cdot\hat{\theta}\rangle,
\label{pl2}
\end{equation}

\noindent

where $\langle x\rangle\equiv x\Theta(x)$, being $\Theta(x)$ the Heaviside
step function.
Now, the flow rule results from the substitution of Eqs.(\ref{pl1}) 
and (\ref{pl2}) into Eq. (\ref{eq:plastic}):

\begin{equation}
d\tilde{\epsilon}^p(\theta)=J(\theta)d\tilde{\sigma}
=\frac{\langle\hat{\phi}\cdot d\tilde{\sigma} \rangle}{h}\hat{\psi}.
\label{eq:flow rule}
\end{equation}

Although we have neither introduced yield functions nor plastic 
potentials, we recover the same structure of the plastic deformation 
obtained from the Drucker-Prager analysis \cite{drucker52}. 
This result suggests the possibility to measure such surfaces directly 
from the envelope responses without need of an a-priori hypothesis 
about these surfaces. The next step is to verify the validity of the 
Drucker-Prager normality postulate, which states that the yield function 
must coincide with the plastic potential function \cite{drucker52}.

\subsection{Normality postulate}

\begin{figure}[b]
 \begin{center}
 \epsfig{file=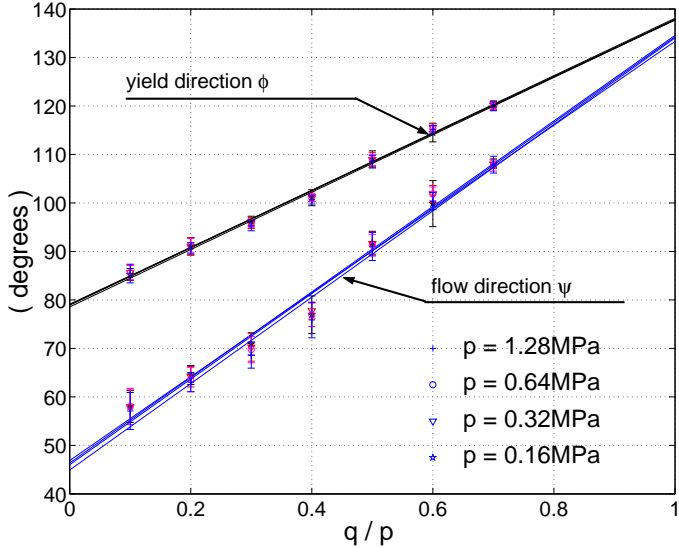,width=0.85\linewidth,angle=-90}
 \end{center}
 \caption{The flow direction and the yield direction of the plastic response. Solid lines represent a linear fit. }
 \label{phipsi}
\end{figure}

The Drucker normality postulate was established to describe the plasticity 
in metals \cite{drucker52}. The question naturally arises as to its 
validity for the plastic deformation for soils. With this aim, the 
yield direction and the flow direction have been calculated for different 
stress states. The results prove that both angles are quite different, 
as shown in Fig. \ref{phipsi}. A large amount of experimental evidence 
has also indicated a clear deviation from Drucker's normality postulate 
\cite{tatsouka74}.

It is not surprising that the Drucker postulate, which has been 
established for metal plasticity, is not fulfilled in the deformation 
of granular materials. Indeed, the principal mechanism of plasticity 
in granular materials is the rearrangement of the grains by the 
sliding contacts. This is not the case of micro-structural changes 
in the metals, where there is no frictional resistance \cite{hill58}. 
On the other hand, the sliding between the grains can be well handled 
in the discrete element formulation, which more adequately describes 
the soil deformation.

The fact that the Drucker postulate is not fulfilled in the deformation 
of the granular materials has led to the so-called non-associated theory 
of plasticity \cite{vermeer84}. This theory introduces a yield surface 
defining the yield directions and a plastic potential function, which 
defines the direction of the plastic strain. 
Both, yield surfaces and plastic potential function can be calculated from 
the yield and flow direction, which in turn are calculated from the strain 
envelope response using Eqs. (\ref{yd}) and (\ref{fd}). According to 
Fig. \ref{phipsi}, they follow approximately a linear dependence with 
the stress ratio $q/p$:

\begin{eqnarray}
\phi &=& \phi_0 +\phi'_0\frac{q}{p},\nonumber\\
\psi &=& \psi_0 +\psi'_0\frac{q}{p}.
\label{eq:stress-dilatancy}
\end{eqnarray}

The four parameters $\psi_0 =46^o \pm 0.75^o$, $\psi'_0 = 88.3^o \pm 0.6^o$, 
$\phi_0 = 78.9^o \pm 0.2^o$ and $\phi'_0 = 59.1^o \pm 0.4^o$ 
are obtained from a linear fit of the data. 
This linear dependence with the stress ratio has been shown to fit well 
with the experimental data in triaxial \cite{rowe62} and biaxial 
\cite{stroud71} tests on sand. In fact, this implies that 
the plastic potential function and the yield surfaces have the 
same shape, independent on the stress level. This is a basic assumption 
from several constitutive models \cite{roscoe68,nova79}.

From Eq. (\ref{eq:stress-dilatancy}), one can see that there is a transition 
from contractancy to dilatancy around $q/p=0.5$. 
This transition is typically observed in dense sand under biaxial loading 
\cite{nova79}.  A consequence of this linear dependency is that $\psi\ne 0$ 
when $q=0$. This implies the existence of deviatoric plastic strain when 
the sample is initially under isotropic loading conditions, which has also
been  predicted in the original Cam-clay model \cite{roscoe68}.

The existence of deviatoric plastic deformation under extremely small 
loading appears to be in contradiction to the fact that the contact 
network remains isotropic below a certain stress ratio. This matter
has also been discussed by Nova \cite{nova79}, who introduced some 
modifications in the Cam-clay model in order to satisfy the isotropic 
condition \cite{nova79}.  However, we are  going to show from a
micro-mechanical inspection that the orientational distribution of the 
sliding contacts departs from the isotropy for extremely small deviatoric 
loadings. 

\subsection{Plasticity \& sliding contacts}

Under small deformations, the individual grains of a realistic soil 
behave approximately rigidly, and the plastic deformation of the 
assembly is due principally to sliding contacts 
(eventually there is fragmentation of the grains, which is not going 
to be taken into account here). A complete understanding of soil 
plasticity is possible, in principle, by the investigation of the
 micro-mechanical arrangement between the grains. We present here 
some observations about the anisotropy induced by loading in the
subnetwork of the sliding contacts. This investigation will be useful 
to understand some features of plastic deformation. 

The sliding condition at the contacts is given by $|f_t|=\mu f_n$, 
where $f_n$ and $f_t$ are the normal and tangential components of 
the contact force, and $\mu$ is the friction coefficient. 
When the sample is isotropically compressed, we observe a significant 
number of contacts reaching the sliding conditions.  If the sample 
has not been previously sheared, the distribution of the orientation 
of the branch vectors of all the sliding contacts is isotropic.

This isotropy, however, is broken when the sample is subjected to 
the slightest deviatoric strain. In effect, at the very beginning 
of the loading, most of the sliding contacts whose branch vector is 
oriented nearly parallel to the direction of the loading leave the 
sliding condition. The anisotropy of the sliding contacts is 
investigated by introducing the polar function $\Omega^s(\varphi)$, 
where $\Omega^s(\varphi)\Delta\varphi$ is the number of sliding contacts 
per particle whose branch vector is oriented between $\varphi$ and  
$\varphi+\Delta\varphi$. This can be approximated by a truncated 
Fourier expansion:

\begin{equation}
\Omega^s(\varphi) \approx \frac{N_0}{2\pi}\big[c_0 + c_1\cos(2\varphi) + c_2\cos(4\varphi) \big].
\label{eq:fabricp}
\end{equation}

\begin{figure}[t]
  \begin{center}
     \epsfig{file=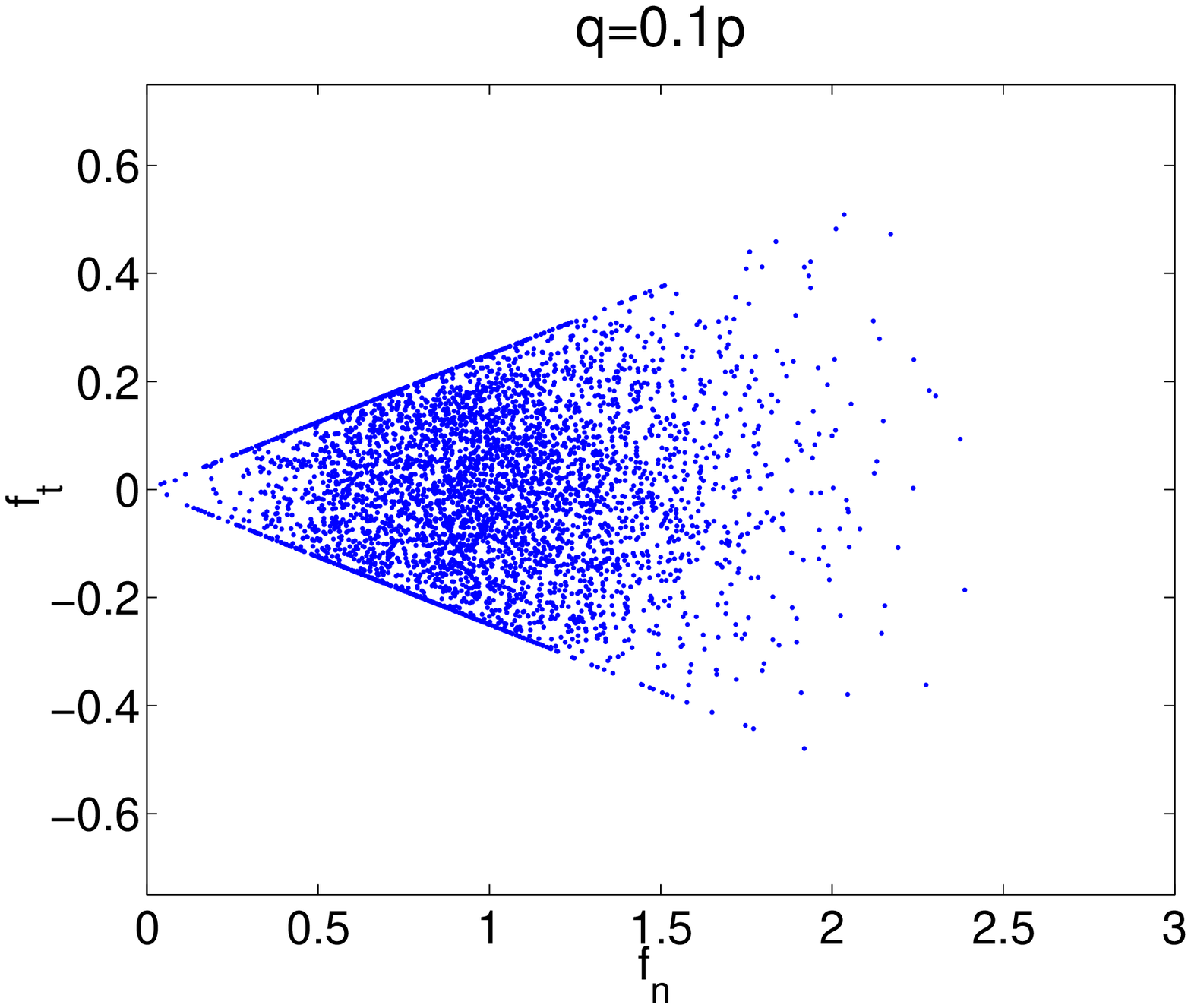,width=0.55\linewidth}
     \epsfig{file=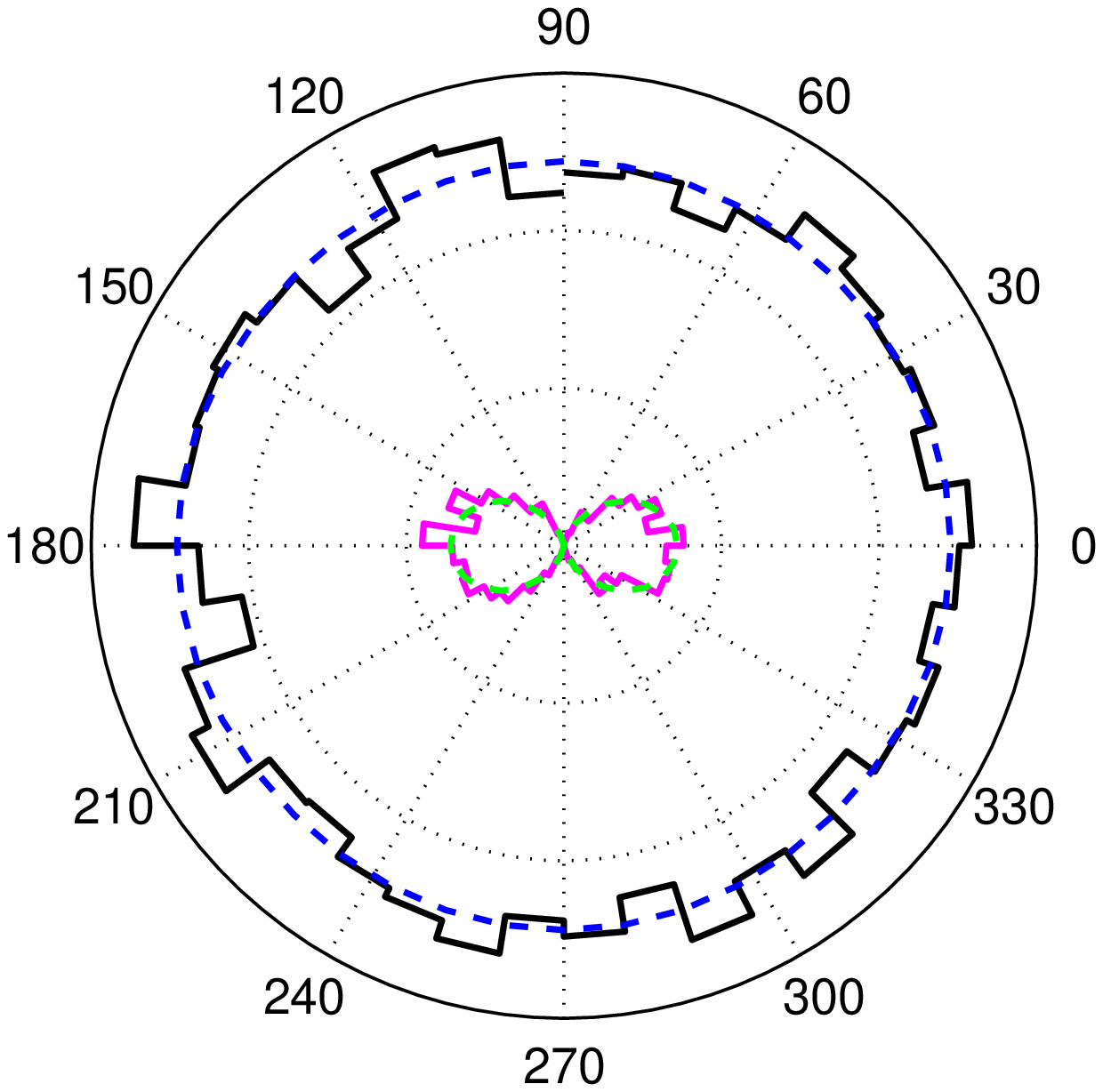,width=0.43\linewidth}
     \epsfig{file=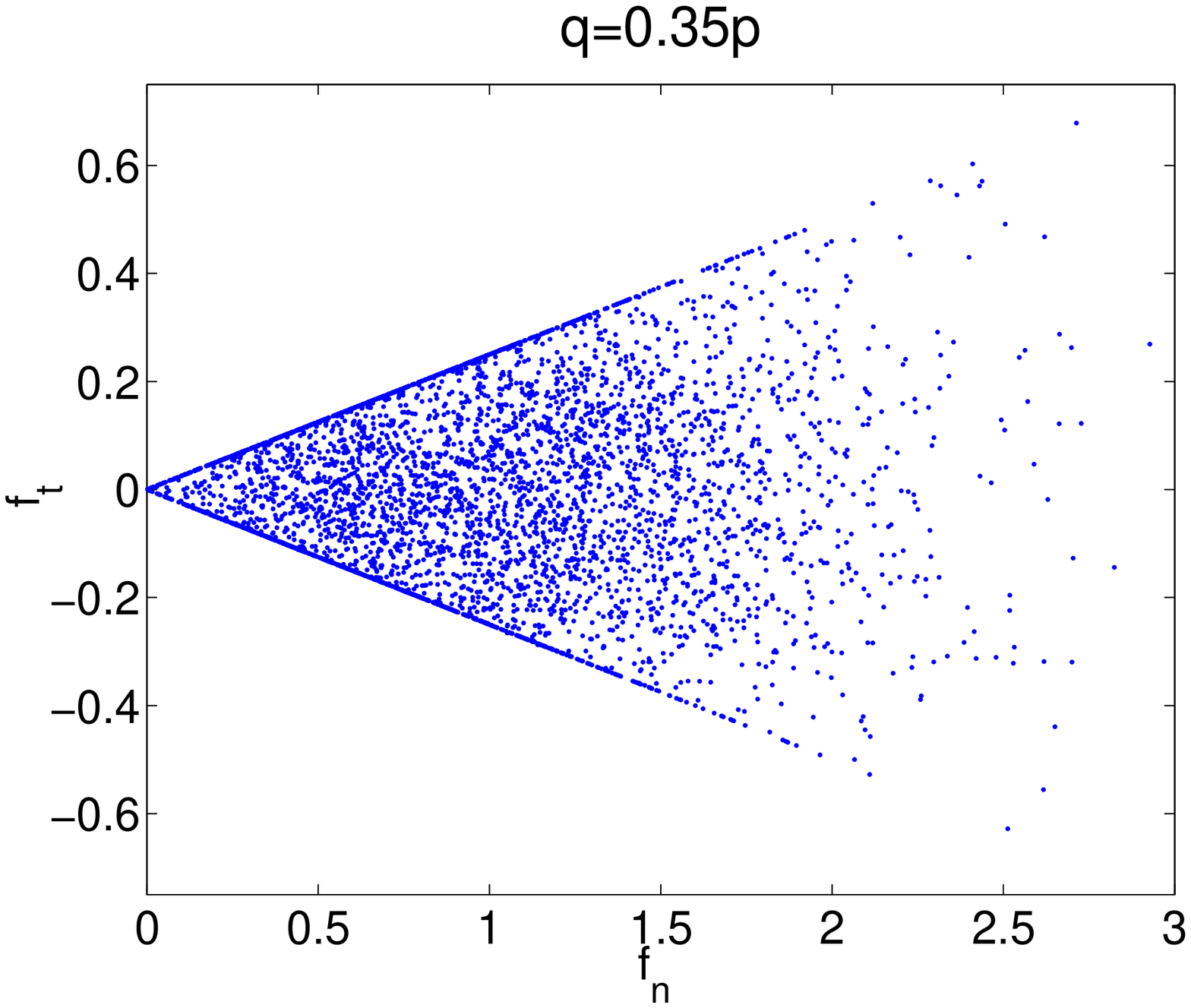,width=0.55\linewidth}
     \epsfig{file=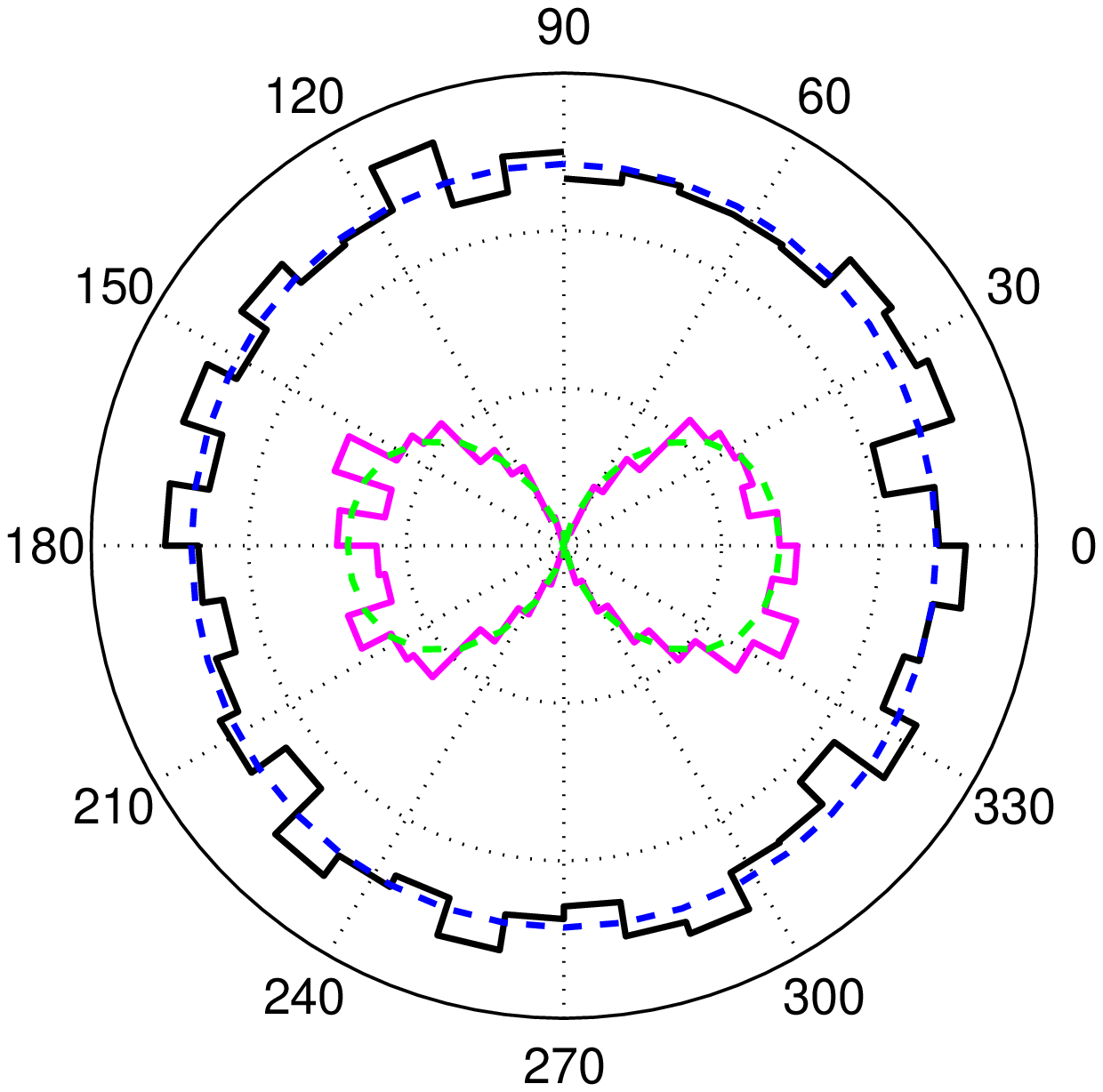,width=0.43\linewidth}
     \epsfig{file=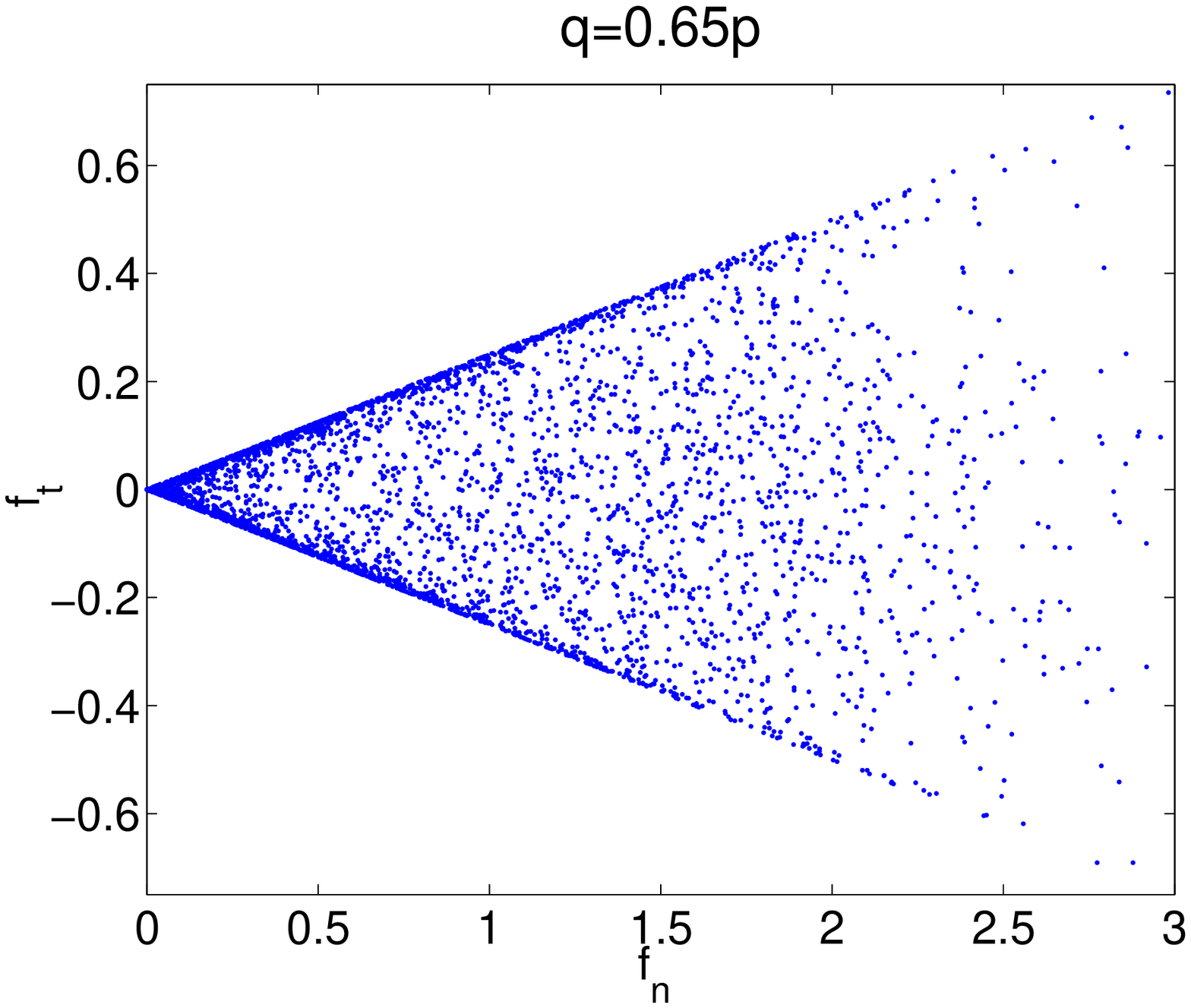,width=0.55\linewidth}
     \epsfig{file=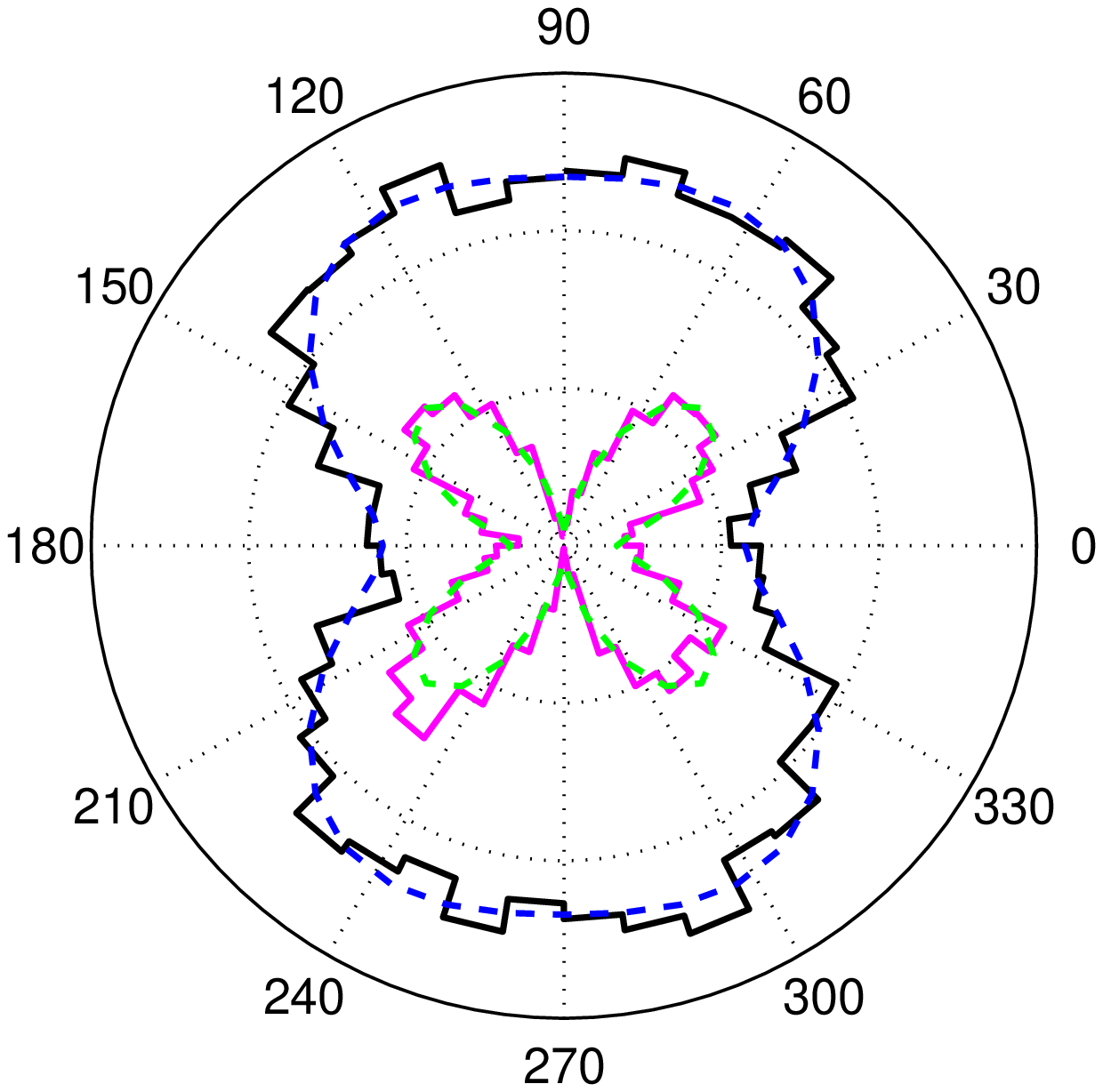,width=0.43\linewidth}
    \caption{Left: force distribution for the stress ratios
          $q/p=0.1,0.35,0.65$. Here $f_t$ and
          $f_n$ are the tangential and normal components of the force. 
          They are normalized by the mean value of $f_n$. Right: orientational
          distribution of the contacts $\Omega(\varphi)$ (outer) and of the sliding 
          contacts $\Omega^s(\varphi)$ (inner). $\varphi$ represents
          the orientation of the branch vector.
          }
    \label{fig:fabric and force}           
  \end{center}
\end{figure}

Fig. \ref{fig:fabric and force} shows the orientational distribution of 
sliding contacts for different stress ratios. For low stress ratios, the 
branch vectors $\vec\ell$
of the sliding contacts are oriented nearly perpendicular
to the loading direction.  Sliding occurs perpendicular to $\vec\ell$, so in this case 
it must be nearly parallel to the loading direction. Then, the plastic 
deformation must be such as $d\epsilon^p_2 \ll d\epsilon^p_1$, 
so Eq.~(\ref{fd}) yields a 
flow direction of $\psi\approx 45^\circ$, in agreement with 
Eq. (\ref{eq:stress-dilatancy}).

Increasing the deviatoric strain results in an increase of the number of 
the sliding contacts. The average of the  orientations of  the branch vectors 
with respect to the load direction decreases with the stress 
ratio, which in turn results in a change of the orientation of the plastic flow. 
Close to the failure, some of the sliding contacts whose branch vectors are
nearly parallel to the loading  direction open, giving rise to a butterfly 
shape distribution, as shown in Fig.  \ref{fig:fabric and force}. In this 
case, the mean value of the orientation of the branch vector with respect 
to the direction of the loading is around $\varphi = 38^o$, which means that 
the sliding between the grains occurs principally around $52^o$ with respect 
to the vertical.  This provides a crude estimate of the ratio between the principal 
components of the plastic deformation as 
$d\epsilon^p_2 \approx -d\epsilon^p_1\tan(52^o)$. According to Eq. (\ref{fd})
this gives an angle of  dilatancy of $\psi=atan2(d\gamma^p,de^p) \approx 97^o$. 
This crude approximation is reasonably close to the angle of dilatancy of 
$103.4^o$ calculated from  Eq. (\ref{eq:stress-dilatancy}). 

A fairly close correlation between the orientation of the sliding contacts 
and the angle of dilatancy has also been reported by Calvetti et al. \cite{calvetti02} 
using molecular dynamic simulations in triaxial tests. This correlation suggests 
that the plastic deformation of soils can be micro-mechanically described by 
the introduction of the fabric constants $c_i$ of the equation (\ref{eq:fabricp})
in the constitutive equations. This investigation would lead to some extensions 
of the fabric tensor capturing the non-associativity of plastic deformation.

\begin{figure}[b]
 \begin{center}
 \epsfig{file=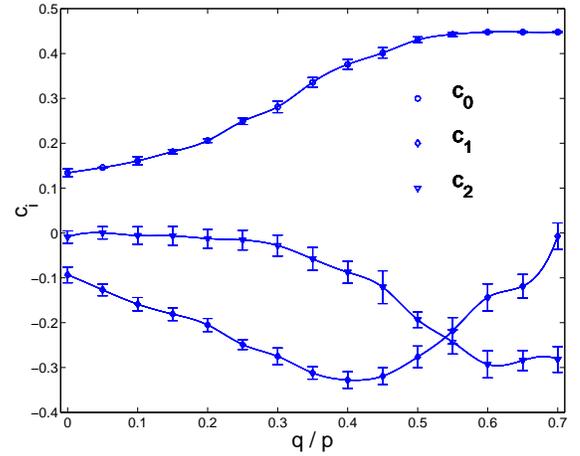,width=0.7\linewidth,angle=-90}
 \end{center}
 \caption{Fabric coefficients of the distribution of the contact normal vectors.
          They are defined in Eq. (\ref{eq:fabricp}).}
 \label{fig:fourier2}
\end{figure}

\subsection{Plastic modulus}

The plastic modulus $h$ defined in  Eq. (\ref{eq:hardening}) is related to 
the incremental plastic strain in the same way as the Young modulus is related 
to the incremental elastic strain. Thus, just as we related the Young modulus 
to the average coordination number of the polygons, it is reasonable to 
connect  $h$ to the fraction of sliding contacts $n_s=N_s/N$. Here $N$ 
and $N_s$ are the total number of contacts and the number of sliding contacts. 

\begin{figure}[b]
  \begin{center}
    \epsfig{file=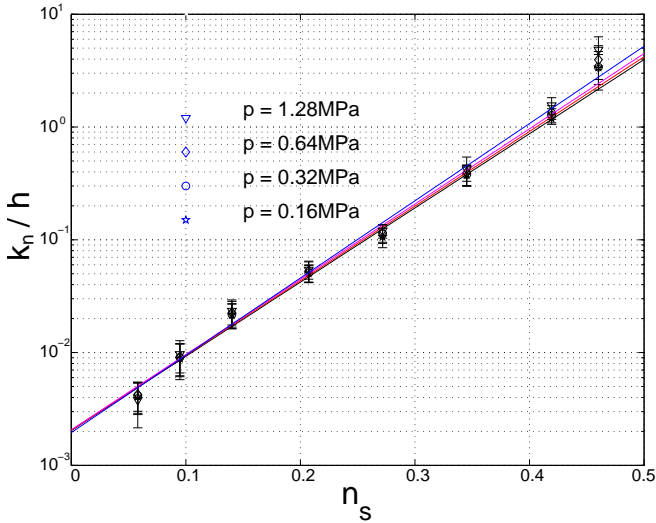,width=\linewidth}
    \caption{Inverse hardening modulus $h$ (MPa) versus fraction of sliding contacts $n_s$.
The values are taken from $q=0,0.1p,...0.7p$ with  different pressures. The lowest 
value of $n_s$ corresponds to $q=0$.}
    \label{fig:h vs ns}
  \end{center}
\end{figure}

Fig. \ref{fig:h vs ns} shows the relation between the hardening and the 
fraction of the sliding contacts.  The results can be fitted to an exponential relation

\begin{equation}
  \label{eq:h vs ns}
    h = h_o\exp(-n_s/n_0)  
\end{equation}
 
Where $h_0=80.0 \pm 0.4 GPa$ and $n_0=0.066\pm0.002$. This exponential 
dependence contrasts with the linear relation between the Young modulus and the 
number of contacts obtained in Sec. \ref{elastic}. From this comparison, it 
follows that when the number of  contacts is such that $n_s>n_0$, the deformation is 
not homogeneous, but is  principally concentrated more and more around the sliding 
contacts as their number increases.

The above results suggest that it is possible to establish a dependency of the flow 
rule on the anisotropy of the subnetwork of the sliding contacts.
This relation is more appropriate than just an explicit relation between the flow rule 
and the stress, which  probes to be only valid in the case of monotonic loading 
\cite{rowe62}. Since the stress can be expressed in terms of micro-mechanical 
variables, branch vectors and contact forces, the identification of those internal 
variables measuring anisotropy and force distribution would provide a more general
description of the dependence of the flow rule on the history of the deformation. 

\section{Concluding remarks}
\label{conclusion}

The effect of the anisotropy of the contact network on the elasto-plastic response 
of a Voronoi tessellated sample of polygons  has been investigated. 
The most salient aspects of this anisotropy are summarized as follows:

\begin{itemize}

\item The incremental elastic response has a centered ellipse as an envelope 
response. Below the stress ratio $q/p<0.4$, this response can be described by 
the two material parameters of Hooke's law of elasticity: the Young modulus 
and the Poisson ratio. Above this stress ratio there is a dependence of the 
stiffness on the stress ratio, which can be connected to the anisotropy 
induced in the contact network during loading. We should state that this 
result might be dependent on the  preparation procedure. In particular,
samples with void ratio different from zero show a smooth transition to the 
anisotropy, which requires further studies.

\item The plastic envelope responses lie almost on the straight line defining 
the plastic flow direction $\psi$. The yield direction $\psi$ and the plastic
modulus $h$ have also been calculated directly from the plastic response. 
The flow direction and yield direction depend on the stress ratio.
In particular, the plastic flow for zero stress ratio has a non-zero 
deviatoric component suggesting an anisotropy induced for extremely small 
deviatoric strains. We found that this effect comes from the fact that the 
sliding contacts depart from anisotropy when the sample is subjected to
the smallest deviatoric deformations. We have also shown a correlation
between the mean orientation of the sliding contacts and the flow direction
of the plastic deformations.

\item In the investigation of the connection between the plastic deformation 
and the number of sliding contacts, we found that the plastic modulus $h$ 
decays exponentially as the fraction of sliding contacts increases. This 
contrasts with the linear decrease of the Young modulus $E$ with the increase 
of the number of open contacts, suggesting that the deformation of the 
granular assembly is concentrated around the sliding contacts.

\end{itemize}

Since the mechanical response of the granular sample is represented by a 
collective response of all the contacts, it is expected that the constitutive 
relation can be completely characterized by the inclusion of some internal 
variables, containing the information about the micro-structural arrangements 
between the grains. We have introduced some internal variables measuring the 
anisotropy of the contact force network. The fabric coefficients $a_i$, measuring 
the anisotropy of the network of all the contacts, prove to be connected with 
the anisotropic stiffness. On the other hand, the fabric coefficients $c_i$, 
measuring the anisotropy of the sliding contacts, are related to the
plasticity features of the granular materials. 

Future work should be oriented towards the elaboration of a theoretical framework 
connecting the constitutive relation to these fabric coefficients. To provide 
a complete micro-mechanically based description of the elasto-plastic features, 
the evolution equations of these internal variables must be included in this 
formalism. This theory would be an extension of the ideas which have been 
proposed to relate the fabric tensor to the constitutive relation of 
granular materials \cite{rothenburg88,thornton86,laetzel02}.

\section*{Acknowledgments}

We thank F. Darve, K. Bagi and F. Calvetti for helpful discussions
and acknowledge the support of the {\it Deutsche Forschungsgemeinschaft\/} 
within the research group  {\it Modellierung koh\"asiver Reibungsmaterialen\/}
and the European DIGA project HPRN-CT-2002-00220.



\end{multicols}

\end{document}